\documentstyle[sprocl,epsfig]{article}

\def\Journal#1#2#3#4{{#1} {\bf #2}, #3 (#4)}

\def\NPB{{\em Nucl. Phys.} B}
\def\NPBP{{\em Nucl. Phys.} B (Proc. Suppl.)}

\def\PRL{\em Phys. Rev. Lett.}
\def\PRD{{\em Phys. Rev.} D}

\def\be{\begin{equation}}
\def\ee{\end{equation}}
\def\bea{\begin{eqnarray}}
\def\eea{\end{eqnarray}}
\begin{document}

\title{THE PHASE DIAGRAM OF 2 FLAVOUR QCD WITH IMPROVED ACTIONS}

\author{M. OEVERS}

\address{Department of Physics and Astronomy, University of Glasgow,\\
         Glasgow G12 8QQ , UK} 

\author{F. KARSCH, E. LAERMANN, P. SCHMIDT}

\address{Department of Physics, University of Bielefeld,\\
Universit\"atsstr.  25, D-33615 Bielefeld, Germany}


\maketitle\abstracts{It has been proposed, that the chiral continuum limit of 2-flavour QCD
                     with Wilson fermions is brought about by a phase in which flavour and parity
                     symmetry are broken spontaneously at finite lattice spacing. At finite 
                     temperature this phase should retract from the weak coupling limit to form 5
                     cusps. This scenario is studied with tree level Symanzik improved actions for
                     both gauge and fermion fields on lattices of size $8^3\times 4$ 
                     and $12^2\times 24\times 4$.}
\section{Introduction and Motivation}
The study of the finite temperature phase diagram of 2-flavour QCD with
Wilson fermions has revealed a rather intricate picture.
This picture is based on the idea of spontaneous breakdown of parity
and flavour symmetry \cite{AokiI} and has been investigated analytically as well 
as numerically \cite{AokiII}. The main features of this phase diagram are: 
\begin{itemize}
  \item the critical line $\kappa_c(\beta)$ defined by a vanishing
    pion screening mass for finite  temporal  lattice size marks the
    phase boundary with a phase of spontaneously broken parity and flavour symmetry. 
  \item for large enough $N_\tau$ five cusps moving towards weak coupling should develop,
    separating the 5 sets of doublers.
  \item the finite temperature phase transition line 
    $\kappa_t(\beta)$ should not cross the critical line, 
    but run past it towards larger values of the hopping parameter,
    presumably \cite{Creutz} turning back toward weak coupling as one crosses the $m_q=0$ line.
\end{itemize}
Because the Wilson term breaks all chiral symmetries of the naive fermion action, the critical line
above cannot consistently be interpreted as the chiral limit of QCD at any finite lattice spacing.
It has recently become clear \cite{SingSharpe} that two scenarios are possible. Either there exists
a second order phase transition to an Aoki phase (of width $a^3$) in which parity and flavour 
symmetry are broken and along its phase boundary the pion mass vanishes or the system exhibits 
a first order transition along the line of vanishing quark mass and the pion does not becomes 
massless. The 
analysis also suggests that which scenario is realized can change as one varies the action. 
One therefore has to check, that the phase diagram with improved Wilson fermions does exhibit an
Aoki phase.
As the Aoki phase retracts from the weak coupling limit as the temperature increases, it naturally 
explains the absence of any non-analyticities across the $m_q=0$ line in the high temperature 
phase.
In order to separate the high temperature side from the low temperature side of the phase
diagram, the finite temperature transition line should run past the cusp of the 
Aoki phase and continue toward larger $\kappa$ values.
The thermal line of the deconfinement phase transition therefore crosses the 
$m_q=0$ line and should bounce back towards weaker coupling due to a symmetry under the change of 
sign of the mass term in the continuum theory \cite{Creutz}. If the thermal line does not touch 
the tip of the cusp, there would be room for the second scenario mentioned above.
\section{The Simulation}
We have simulated 2 flavours of Wilson fermions on lattices of size $8^3\times 4$ and 
$12^2\times 24\times4$. We have employed tree level Symanzik improvement which for the fermion 
sector amounts to adding the so called clover term with coefficient one and for the gauge fields
to adding a $2\times 1$ loop. We have first mapped out the phase diagram on the smaller lattice
and have then corroborated our results on the larger lattice for two $\beta$ values.
\begin{figure}[t]
   \begin{center}
   \epsfig{file=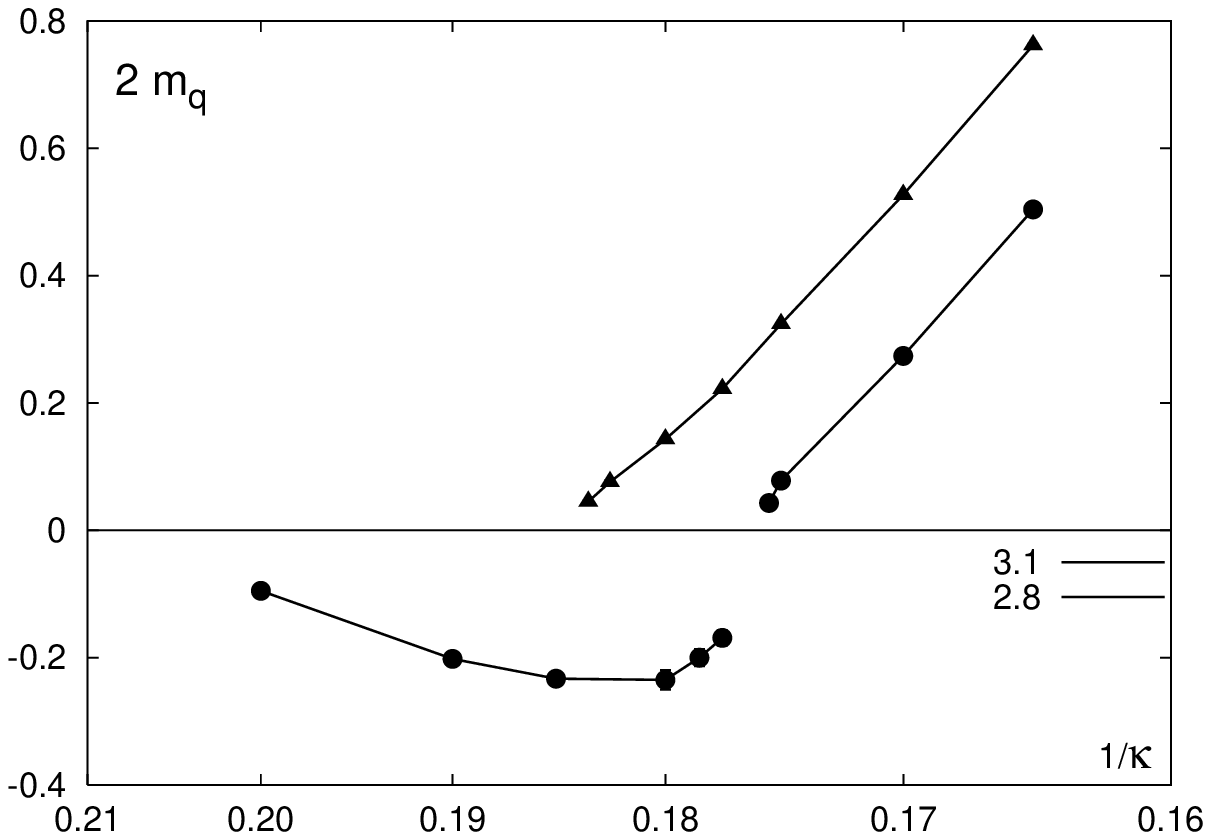,width=5.9cm}
   \epsfig{file=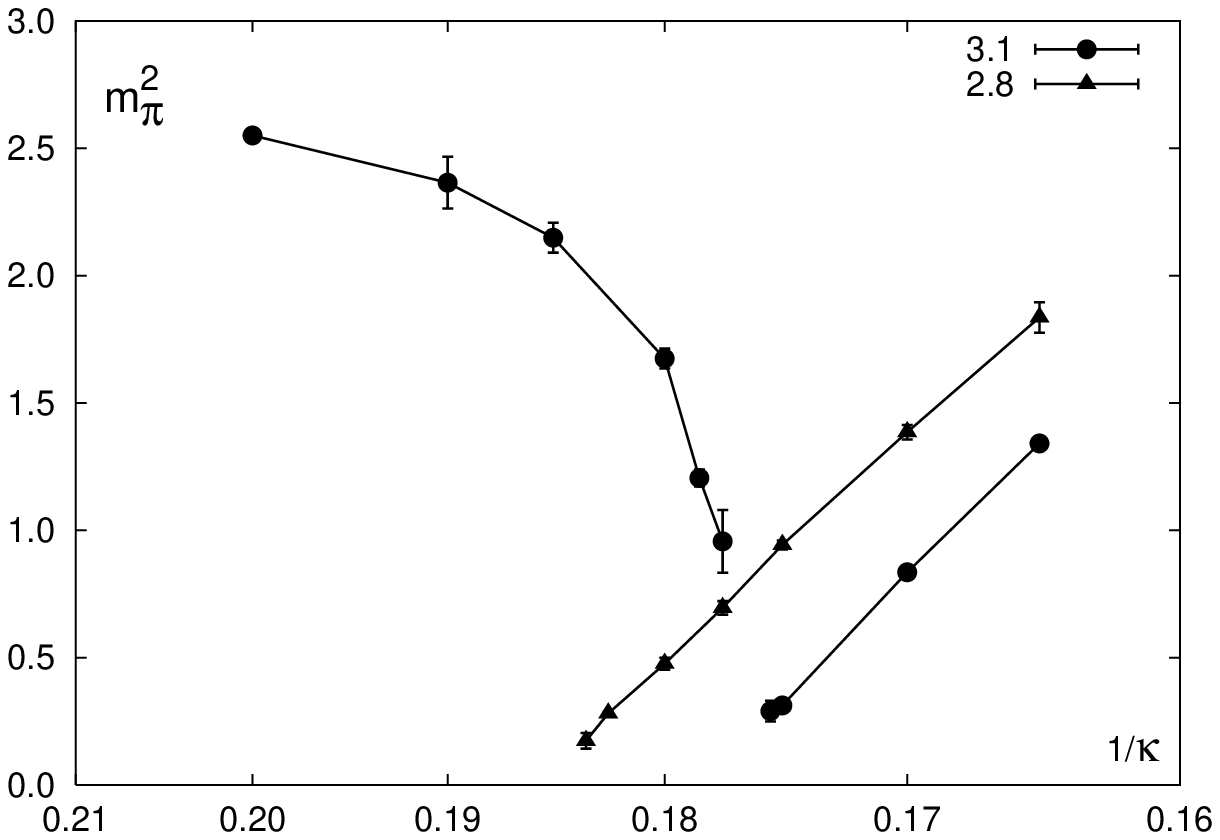,width=5.9cm}
   \caption{ \label{f:mqpi} $m_q$(left) and $m_\pi^2$(right) as a function of $1/\kappa$ on 
            the $12^2\times 24\times 4$ lattice.}
   \end{center}
\end{figure}
\section{Quark mass and Pion screening mass }
Our results for the quark and pion screening mass are shown in figure (\ref{f:mqpi}).
The current quark mass is defined via the axial Ward identity \cite{Bochi}:
\begin{equation}
  2m_q \equiv \frac{\nabla_\mu \langle 0 |  A^\mu | \pi \rangle}
  {\langle 0 | P | \pi \rangle}
\end{equation}
For $\beta=2.8$ we find some curvature for the quark mass as a
function of $1/\kappa$, though a linear fit produces a reasonable
$\chi^2$. For $\beta=3.1$ we have also explored a region of hopping
parameters where the quark mass becomes negative. There we find a
rather peculiar behaviour. The quark mass first decreases as one lowers
$\kappa$ towards  $\kappa_c$ and only rises again very close to $\kappa_c$. This results in  
different slopes of $m_q$ as a function of $1/\kappa$ for positive and negative quark masses, 
which is in contrast to simulations with unimproved Wilson fermions that have
shown the same behaviour for  positive and negative quark
masses~\cite{AokiII}. 
We have also accurately measured the pion screening mass. For
$\beta=2.8$ the decrease of the pion mass is consistent with a linear
behaviour $m_\pi^2 \propto 1/\kappa - 1/\kappa_c$ down to small values
of $m_\pi^2$. This also applies for $\beta=3.1$ for $\kappa$ values
that correspond to positive quark masses.
For negative quark masses the behaviour is quite different and
inconsistent with a linear behaviour for the points  measured. We will
however argue below that for $\beta=3.1$ the pion mass does not go to zero as
the quark mass goes to zero, because one crosses the finite
temperature transition line before the quark mass becomes zero. 
\section{Polyakov loop}
\begin{figure}[t]
   \begin{center}
   \epsfig{file=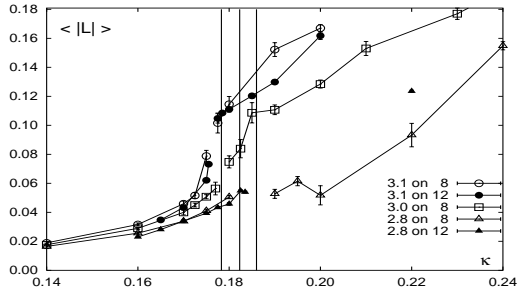,width=7cm,height=4cm}
   \caption{ \label{f:poly} Polyakov loop as a function of $\kappa$, open symbols for the 
            smaller lattice, filled symbols for the larger lattice.}
   \end{center}
\end{figure}
Figure (\ref{f:poly}) shows our results for the Polyakov loop as a function of
$\kappa$ including data from both lattices. The vertical lines
indicate the position of the extrapolated $\kappa_c$ from the pion screening mass, except for
$\beta=3.0$ where the pion norm was used. This extrapolated $\kappa_c$ decreases with increasing 
$\beta$. 
For $\beta=2.8$ all our data for the pion mass
lie in the confined region. We therefore conclude that the pion mass
vanishes as the quark mass goes to zero, i.e. for $\beta=2.8$ we hit
the Aoki phase as we increase $\kappa$. For $\beta=3.0$ and $3.1$ this
is no longer so clear. The Polyakov loop already  shows a high
temperature behaviour where the extrapolated pion mass would be
small. For $\beta=3.0$ the situation is less prominent and one could
still argue that the pion becomes massless, but since the Polyakov
loop is in the high temperature phase immediately after one crosses
$\kappa_c$,  the finite temperature line and the line of vanishing
quark mass come very close together. For $\beta=3.1$ it becomes
evident, that one crosses the finite temperature transition line
before the line of vanishing quark mass. The pion will therefore not
become massless as the quark mass vanishes.  Because the Polyakov loop
continues to rise past the $m_q=0$ line, one can exclude that the
finite temperature line bounces back towards weaker coupling
immediately.   

%\section{Locating the critical line}
%To locate the critical hopping parameter $\kappa_c$ we have performed
%the following fits to our data. We have fitted $m_\pi^2$ linearly and
%$m_q$  quadratically as a function of $1/\kappa$. Since the pion norm
%near $\kappa_c$  is expected to behave like $\Pi \approx 1/m_{\pi}^2$,
%we have also extracted a value for $\kappa_c$ from this
%observable. Since the data show quite some nonlinear behaviour as a
%function of $1/\kappa$, we have used a quadratic fit ansatz. As the
%results from the different fits turned out to be consistent, we took
%confidence in using our data for the pion norm from our simulation at
%$\beta=3.0$ on the smaller lattice to get an estimate of $\kappa_c$
%for $\beta=3.0$. Our results are summarized in the following table:
%\begin{center}
%  \begin{tabular}[t]{|c||c|c|c|} \hline
%    $\kappa_c$ & $ \beta= $2.8 & $ \beta= $3.0 & $ \beta= $3.1 \\ \hline
%    $m_\pi$    & 0.1860(2)     &               & 0.1783(2)     \\ \hline
%    $m_q$      & 0.1853(3)     &               & 0.1770(3)     \\ \hline
%    $\Pi$      & 0.1859(3)     & 0.1823(10)    & 0.1800(5)     \\ \hline
%  \end{tabular}
%\end{center}
\section{Chiral order parameter}
\begin{figure}[t]
   \begin{center}
   \epsfig{file=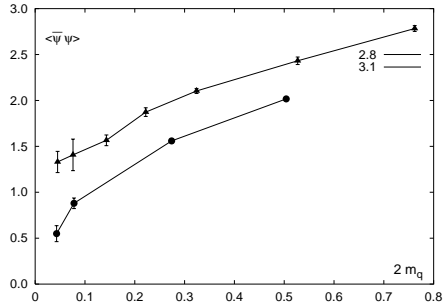,width=6cm}
   \caption{ \label{f:psi} Chiral order parameter as a function of the quark mass on 
              the $12^2\times 24\times 4$ lattice.}
   \end{center}
\end{figure}
Because of the explicit breaking of chiral symmetry by the Wilson
action, one has to define a properly subtracted chiral order parameter to
obtain the correct continuum limit. Using axial Ward identities the
order parameter is defined as follows \cite{Bochi}: 
\begin{equation}
  \langle\bar{\psi} \psi\rangle_{sub} = 2m_q \cdot Z \cdot 
  \sum_{x}\langle\pi(x)\pi(0)\rangle
\end{equation}
Here $Z$ is a renormalisation factor for which we take its tree level
value  $Z=(2\kappa)^2$. The sum over the pion correlation function is
just the pion norm. Our results are shown in figure (\ref{f:psi}).
For $\beta=2.8$ the data extrapolate to a finite
intercept at $m_q=0$.  For $\beta=3.1$ the data show more curvature
and we expect  $\langle\bar{\psi} \psi\rangle_{sub}$ to go to zero 
for vanishing quark mass.
\section*{Acknowledgments}
Work in part supported by NATO grant no. CRG940451 and by
the  European Community TMR Programs TRACS and ERBFMRX-CT97-0122

\end{document}